# Amorphous-like Density of Gap States in Single Crystal Pentacene

D. V. Lang[1], X. Chi[2], T. Siegrist[3], A. M. Sergent[3], and A. P. Ramirez[3]

[1]Los Alamos National Laboratory, Los Alamos, NM 87545
[2]Columbia University, New York, NY
[3]Bell Laboratories, Lucent Technologies, 600 Mountain Avenue, Murray Hill, NJ 07974

We show that optical and electrical measurements on pentacene single crystals can be used to extract the density of states in the HOMO-LUMO bandgap. It is found that these highly purified crystals possess band tails broader than those typically observed in inorganic amorphous solids. Results on field effect transistors (FETs) fabricated from similar crystals are also compared. The FET data imply that the gap state density is much larger within 5-10 nm of the gate dielectric. These results are discussed in terms of both crystal phase domains and structural disorder mechanisms.

Organic semiconductors, such as pentacene, are promising for applications requiring large-area coverage, flexibility, low-temperature processing, and low cost [1-4]. However, there is evidence that defects and/or impurities limit the maximum device performance. The effective mobility of holes in pentacene increases with increasing crystal perfection in thin-film transistors (TFTs) [4]. Völkel et al. [5] demonstrated that their polycrystalline pentacene TFTs were limited by tails of localized states with a higher concentration 5-10nm from the gate dielectric than in the bulk of the film. Our previous results on single-crystal pentacene field-effect transistors (FETs) indicate that the effective mobility of holes is limited by localized states near the valence band [6]. In addition, Karl and co-workers [7] showed convincingly that maximum time-of-flight mobility in various organic single crystals can only be obtained in ultra-purified material.

In this paper we will show that purified pentacene single crystals have significant exponential band tails that are broader than those observed in inorganic amorphous material. We also show that FETs fabricated from similar pentacene crystals have broad tails of valence band states near the gate dielectric, with a concentration at least an order of magnitude larger than in the bulk. This is an important result, since it shows that purified single crystals of pentacene have band tails that are suggestive of disordered material. It is also important because it implies that the performance of pentacene FETs is not limited by the properties of the underlying material but by localized defects and/or impurities near the interface with the gate dielectric.

Band tails of localized states are typically observed in optical absorption or photoconductivity spectra. Inorganic crystals [8, 9] have narrow, temperature-dependent, band tails (Urbach tails) given by $\sim\exp(-(E_g-E)/nkT)$, where $E_g$ is the band gap, $E$ is the photon energy, k is Boltzman's constant, T is the absolute temperature, and n ~ 1. Amorphous inorganic materials [9] typically have broader band tails, given by $\sim\exp(-\beta(E_g-E))$, with the slope parameter $\beta$ in the range 10 to 25 eV$^{-1}$ at 300K. These band tails are caused by disorder and are temperature independent below the glass transition temperature. Band tails in organic materials are usually not measurable by these methods because the intra-molecular singlet exciton dominates this part of the spectrum [10]. In pentacene, however, it is possible to observe the single-particle gap states in the region of the singlet exciton absorption by noting that the generation of photo-induced carriers by singlet excitons is an extrinsic process [10-12]. The singlet exciton dominates the optical absorption spectrum of pentacene [11, 13-15], but these excitons decay with a rate of $1.3 \times 10^{13}$ sec$^{-1}$ into pairs of triplet excitons [16]. The mobile triplets decay nonradiatively by interacting with paramagnetic oxygen molecules [17], free carriers[17], or trapped carriers [11, 12]. Photoconductivity in this spectral range is thus due both to the excitation of single-particle gap states as well as the release of trapped carriers by triplet excitons. The relative efficiency of the exciton-mediated photoconduction process depends on impurities and/or defects and can be greatly reduced under appropriate conditions.

We have recently discovered a metastable hole trap in pentacene that is rapidly quenched by the capture of electrons generated by photon absorption above the HOMO-LUMO bandgap [18]. We report here the spectral response of this photo-quenching rate and show that this process is only weakly dependent on the decay of triplet excitons, thereby allowing us to extract the underlying band tails of the pentacene single crystal.

The pentacene crystals were grown by physical vapor transport as described previously [6]. Typical crystals used for the experiments were 10-30mm$^2$ in area and 25-50μm thick. To evaluate crystal quality, rocking curves of a sample from the same batch as the transport sample were obtained on a custom four-circle diffractometer with monochromated Cu-K$\alpha$ radiation. Rocking curves of (002) and (004) reflections were evaluated, and the widths indicated mosaic spreads of the order of 0.2 degrees and better. Due to the intrinsic resolution of the diffractometer further evaluation will require tighter resolution. The contacts were formed by evaporating 5nm of Ti followed by 500nm of Au in an e-beam evaporator through a shadow mask giving an array of 400µm-square metal pads separated by 30µm gaps along the columns and 108µm gaps along the rows. The measurements reported in this paper were made on the 30µm gaps with a specular surface. The contacted crystals were mounted in a temperature-variable vacuum probe station and measured with a Keithley 6517A electrometer. Photo-excitation was achieved with a grating monochromator focused to a 1mm-diameter spot, giving an optical flux ranging from 3 x $10^{15}$ photons/cm$^2$sec to 2 x $10^{12}$ photons/cm$^2$sec at wavelengths of 1000nm and 350nm, respectively. Photoconductivity (10Hz chopping rate) was measured at 293K with 120V (4 x $10^4$ V/cm) applied to the sample. The photo-quenching rate of the metastable hole trap was measured at 291-293K by first polarizing the sample at 600V for 100sec and then measuring the current transient following a bias step to 120V with the photon flux switched on at a preset wavelength. Under these conditions, thermal recovery in the dark is negligible [18]. This sequence was repeated for various wavelengths to obtain the photo-quenching spectrum.

Figure 1 shows the photoconductivity yield (electrons/photon) versus photon energy under the condition outlined above. We also show the photoconductivity yield of Silinsh et al [11] measured at 1 x $10^4$ V/cm on a 0.6-2µm-thick polycrystalline film with semi-transparent Au electrodes in a sandwich configuration. Note that these data agree remarkable well with our data above the 2.2eV HOMO-LUMO gap with no adjustable parameters. Our data for the photo-quenching rate divided by the photon flux is shown in Fig. 1 scaled to fit the photoconductivity data above 2.2eV. The HOMO-LUMO absorption edge in Fig. 1 is fit with the function $(E-E_g)^n/E^2$ with n=2.5 and $E_g$=2.25eV; this power law with n=2 is typical of the optical absorption edge in many amorphous solids.[9] We note that Silinsh et al. [11] originally fit their data to the function $(E-E_g)^{5/2}$ with $E_g$=2.20eV; this power law is typical of the photoconductivity threshold in many organic crystals [10].

The photo-quenching data, along with the data of Silinsh et al. [11], define an exponential band tail with a slope of β = 5.7eV$^{-1}$ plus an exciton feature that is 0.4% times the photoconductivity. We believe this band tail is a good measure of the bulk gap states in pentacene. Both of these samples were purified by two- or three-fold vacuum sublimation before the final growth; therefore, we believe these band tails are most likely due to disorder rather than impurities. The large difference in our sample between photoconductivity yield (due to both electrons and holes) and photo-quenching rate (due to only electrons) implies that the rate for the release of trapped electrons by triplet excitons in pentacene is orders of magnitude less than the rate for the release of holes. The difference between the exciton effects in our photoconductivity data compared to those of Silinsh et al may be due to such factors as the sample geometry, the existence of grain boundaries in the polycrystalline film, or differences in the concentration of paramagnetic oxygen molecules.

We will now extract the density of gap states from our FET data [6] shown in Fig. 2a, which gives the activation energy of the source-drain current at 40V as a function of the gate bias in the temperature range of 200-300K. The density of gap states, N(E), can be obtained from these data by the following argument. Since the concentration of free carriers is much less than the concentration of trapped carriers, the charge induced by the gate bias is primarily a measure of the density of gap states. The charge per unit area induced in the FET by a gate bias $V_g$ is given by $Q=CV_g$, where C is the gate capacitance per unit area. The area density of states, N(E) in states/cm$^2$eV, is given by

$$N(E) = (C/q)(1/(dE/dV_g)), \qquad (1)$$

where q is the electronic charge and E is the energy measured from the valence band edge; E is also the activation energy in Fig. 2a. The gate capacitance for the device in ref. [6] is $C=6\times10^{-9}$ F/cm$^2$. The area density of states in Eqn. 1 can be converted into a volume density by determining the depth over which the carriers are trapped. As pointed out by Völkel et al. [5], the mobile holes in the FET accumulation layer are just a few nm from the interface; therefore, the gap states that control the local Fermi level in the simulation of the accumulation layer must be 5-10 nm of the interface.

Figure 2b gives the volume density of states obtained from the smooth fit to the data in Fig. 2a using Eqn. 1 and assuming a trapping depth of 7.5nm. We note in Fig. 2a that the activation energy data at threshold, $V_g \sim 2.5V$, is not very reliable; therefore, in this region we use the smooth curve to calculate the derivative in Eqn. 1. The best fit to these data is a peak at 0.35eV plus an exponential band tail $N_0\exp(-\beta E)$, with $N_0 = 9\times10^{19}$ cm$^{-3}$eV$^{-1}$ and $\beta = 9.2$ eV$^{-1}$. The density of states functions obtained by Völkel et al. [5] for donor and acceptor defects are shown in Fig. 2b for comparison. We cannot determine from our data whether our defect states are donors or acceptors, but we note the similarity in magnitude to the density-of-states function for acceptors used by Völkel et al. [5], who point out that their slope $\beta = 5$ eV$^{-1}$ remains the same while the magnitude $N_0$ varies somewhat from sample to sample. This similarity between polycrystalline TFTs and a single crystal FET suggests that the density of defects near the gate interface is similar in both cases, despite the very different fabrication processes involved. It is interesting to note that in both cases the effective mobilities are also very similar: 0.3 cm$^2$/Vsec for the single crystal device [6] and 0.3 to 0.6 cm$^2$/Vsec for the polycrystalline TFTs.[5] We also note that the peak at 0.35eV is the same as the Fermi energy region associated with the onset of metastable hole trapping in pentacene.[18]

Figure 3 shows both the optical data from Fig. 1 and the FET data from Fig. 2b on the same density-of-states scale with the energy measured from the top of the valence band. The density-of-states scale is determined by normalizing the peak of the band-to-band photoconductivity yield in electrons per photon (at ~2.8eV) to the molecular density of pentacene (2.9 x 10$^{21}$ cm$^{-3}$) divided by the FWHM of this band (~0.5eV). This photoconduction peak is actually a joint density of states for both the valence band and conduction band, but we believe this normalization is relevant for determining the density of gap states since the single-particle excitations from the valence band tail are also a convolution with the conduction band final states. Note that the band tail obtained from the FET data has a significantly larger concentration than the band tail apparent in the optical data with weak exciton effects. Since the FET data corresponds to traps within 5-10nm of the parylene gate dielectric deposited on the crystal surface, these results imply that such defects or impurities are either introduced by the parylene deposition process or exist near the surface of the crystal before the gate fabrication. In either case, we would expect that device performance could be improved if these band tails were reduced to a level more typical of the pentacene bulk.

The origins of the bulk and interface band tails in pentacene are presently unknown; however, we can consider some possible causes. First, we note that in molecular crystals a significant electronic polarization reduces the band gap of the solid from the larger HOMO-LUMO gap of the molecule [10]. As a result, the local band gap could depend strongly on the local lattice constant, which could vary as a result of inhomogeneous strains or disorder within the crystal. Indeed, calculations have shown that lattice relaxation near a point defect in anthracene could create a 0.28eV trap level and the strain field near a dislocation could give a 0.2eV trap [10, 19]. Tiago et al have shown from accurate *ab initio* calculations that the band gap of pentacene can vary by 0.3eV between the two reported polymorphs of the pentacene crystal structure [15]. In addition, it is well known that thin films of polycrystalline pentacene exhibit a different metastable crystal phase [13, 20]. Thus, one might expect that pentacene solids of one predominant crystal structure could nevertheless have small regions of other polymorphs or defects present that could be responsible for the band tails that we observe. These polymorphic or disordered domains might also be related to metastable trapping by local structural changes resulting from external perturbations, as has been suggested by Knipp et al. [20].

In conclusion, we have shown that optical and electronic measurements of pentacene material and devices, based on both single crystals and polycrystalline thin films, can be interpreted to imply several important general properties of this class of organic semiconductors. First, state-of-the-art purified pentacene single crystals still have sufficient disorder and/or impurities to have substantial band tails, even broader than the band tails observed in inorganic amorphous solids. Because our samples have been highly purified and because such results are also observed for purified pentacene the literature, we favor the intrinsic disorder scenario. Second, FET devices fabricated from pentacene single crystals and polycrystalline thin films can have an order of magnitude more valence-band-tail states within 5-10nm of the gate dielectric than is typical of bulk pentacene in these two forms. This suggests that devices can be substantially improved by reducing these interface-related states to the bulk concentration level or below.

We acknowledge the support of the US Department of Energy under grant # 04SCPE389.

**Figure Captions**

Figure 1. Photocurrent yield (electrons/photon) for single crystal pentacene versus photon energy. The photo-quenching rate of the metastable hole trap is normalized to the photocurrent above 2.2eV. The photocurrent yield data of Silinsh et al is plotted without adjustable parameters. Various fits to the data are described in the text.

Figure 2. (a) FET data of Butko et al with a smooth fit to the data as described in the text. (b) Density of states, N(E), obtained from the FET data in part (a) of this figure using a trapping depth of 7.5nm. The band tails of Völkel et al and the fit to the FET band tail are described in the text.

Figure 3. Photo-quenching rate and Silinsh photocurrent yield normalized to a density of states scale as described in the text. The FET density of state is from Fig. 2(b). The FET and optical band tails are given as a function of energy measured from the top of the valence band. The HOMO-LUMO curve is the photocurrent threshold lineshape function given in Fig.1.

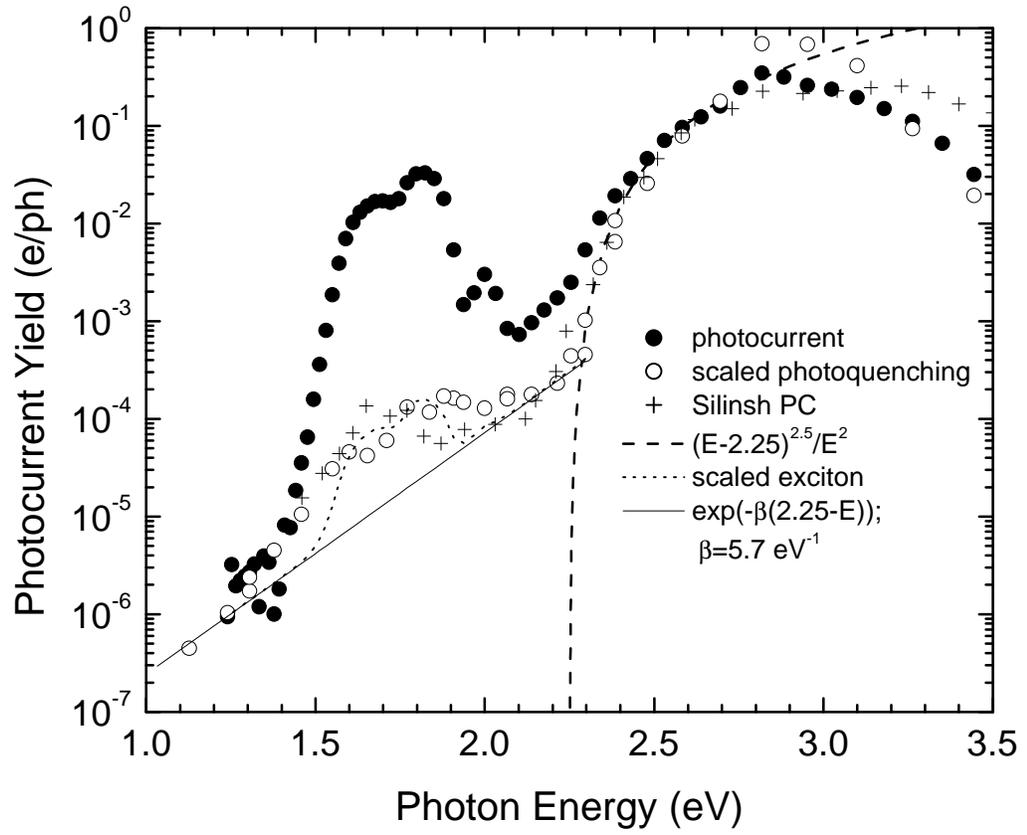

Fig. 1

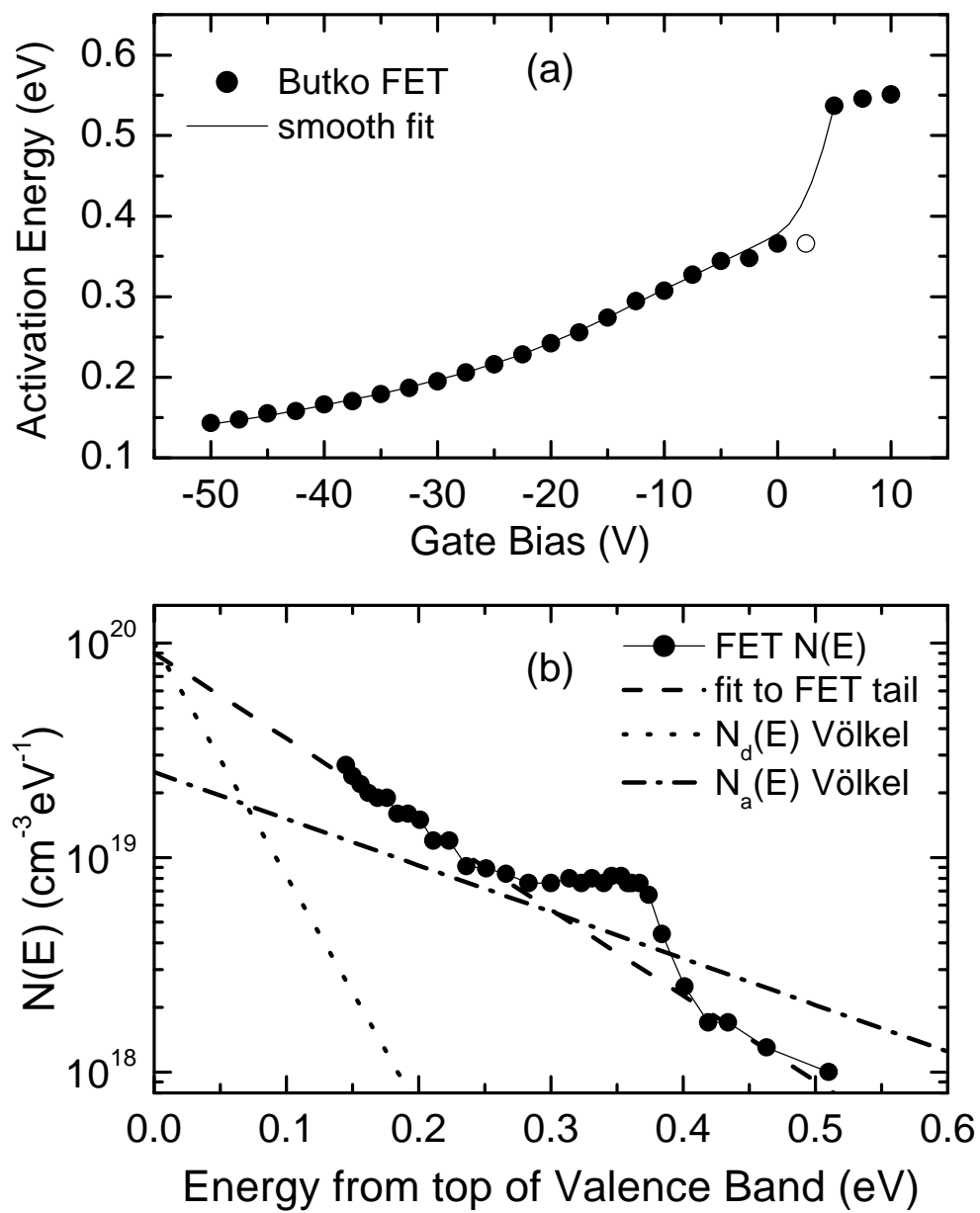

Fig. 2

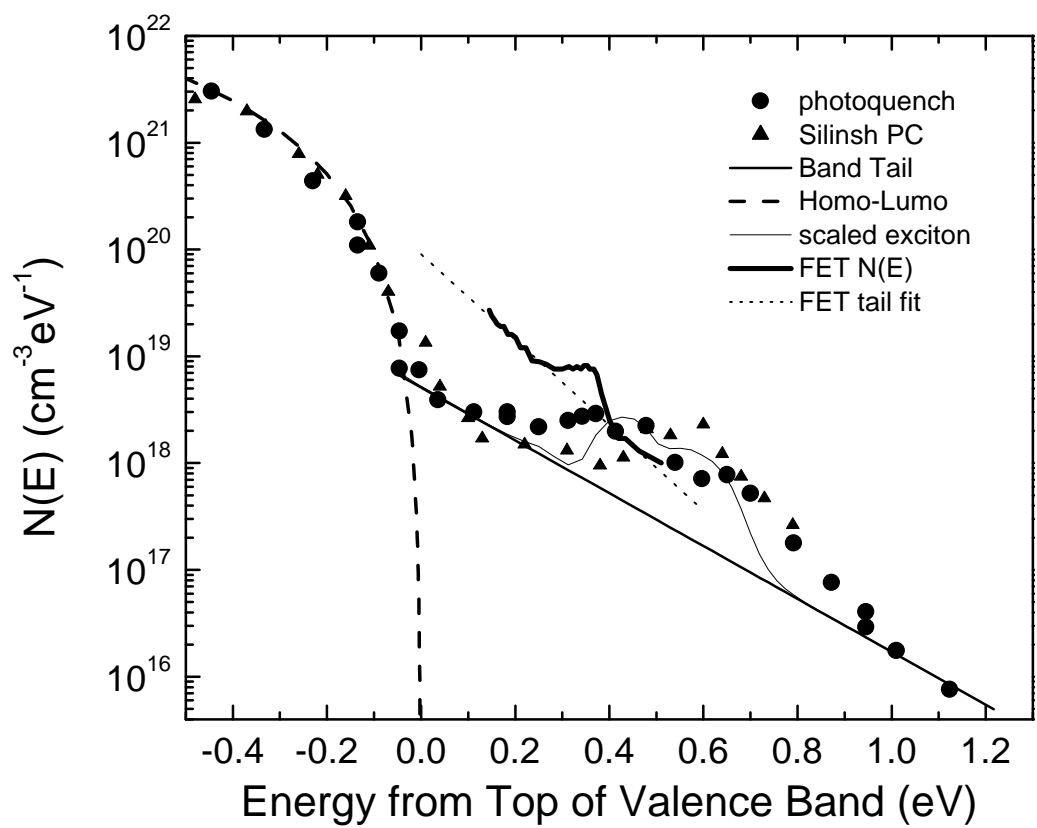

Fig. 3